\documentclass[12pt,letterpaper]{article}
\usepackage{hyperref}
\usepackage{amsmath,amsfonts,amssymb}
\usepackage{xspace}
\usepackage{graphicx}
\numberwithin{equation}{section} 

\def\cN{\mathcal{N}}
\def\bZ{\mathbb{Z}}
\def\GG{\mathcal{G}}
\def\ch{\mathop{\mathrm{ch}}\nolimits}
\def\Tr{\mathop{\mathrm{Tr}}\nolimits}

\begin{document}

\begin{titlepage}
\begin{flushright}
\normalsize
{\tt AEI-2010-048, CALT-68-2783,}\\
{\tt IPMU-10-0053, YITP-10-20}\\

\medskip

April, 2010 
\end{flushright}
\vfil

\bigskip

\begin{center}
\Large Notes on the K3 Surface and the Mathieu group $M_{24}$
\end{center}

\vfil
\medskip

\begin{center}
\def\AA{1}
\def\BB{2}
\def\CC{3}
\def\DD{4}
\def\EE{5}
Tohru Eguchi$^\AA$, \quad
Hirosi Ooguri$^{\BB,\CC,\DD}$ \quad
and \quad 
Yuji Tachikawa$^\EE$

\bigskip
\bigskip
\itshape

$^\AA$ Yukawa Institute for Theoretical Physics, \\
Kyoto University, 
Kyoto 606-8502 Japan

\medskip

$^\BB$ California Institute of Technology, Pasadena, CA 91125, USA\\

\medskip
$^\CC$ Institute for the Physics and Mathematics of the Universe,\\
University of Tokyo, Kashiwa 277-8582, Japan\\

\medskip
$^\DD$ Max-Planck-Institut f\"ur Gravitationsphysik, \\
 D-14476 Potsdam, Germany\\

\medskip

$^\EE$ School of Natural Sciences, Institute for Advanced Study,\\
Princeton, NJ 08540, USA
\end{center}

\vfil
\bigskip

\begin{center}
{\bfseries Abstract}
\end{center}

\bigskip

We point out that the elliptic genus of the K3 surface 
has a natural decomposition in terms of dimensions of
irreducible representations of the largest Mathieu group 
$M_{24}$. The reason is yet a mystery.

\vfill

\end{titlepage}
\addtocounter{section}{1}

Elliptic genus of a complex $D$-dimensional hyperK\"ahler manifold $M$ is defined as 
\begin{equation}
Z_{ell}(\tau;z)=\Tr_{{\cal R}\times {\cal R}} (-1)^{F_L+F_R}q^{L_0}\bar{q}^{\bar{L}_0}e^{4\pi iz J^3_{0,L}}\label{def-elliptic-genus}
\end{equation}  in terms of the two-dimensional supersymmetric sigma model whose target space is $M$ \cite{WittenElliptic}. 
As $M$ is assumed to be hyperK\"ahler, the two-dimensional theory has $\cN=4$ superconformal algebra as its symmetry.
Then  $L_0$ and $\bar{L}_0$ are zero modes of the left- and right-moving Virasoro operators; $J^3_0$ is the zero mode of the 3rd component of the affine $SU(2)$ algebra; $F_L$ and $F_R$ are the left- and right-moving fermion numbers. Trace is taken over the Ramond sector of the theory. 
This elliptic genus is a Jacobi form of weight=0 and index ${D\over 2}$.

Elliptic genus for K3 surface was 
explicitly calculated in \cite{EOTY} and is given by 
\begin{equation}
Z_{ell}(K3)(\tau;z)=8\left[\left({\theta_2(\tau;z)\over \theta_2(\tau;0)}\right)^2
+\left({\theta_3(\tau;z)\over \theta_3(\tau;0)}\right)^2
+\left({\theta_4(\tau;z)\over \theta_4(\tau;0)}\right)^2\right].
\label{K3-elliptic-genus}
\end{equation} 
Here $\theta_i(\tau;z)\,(i=2,3,4)$ are Jacobi theta functions. 
Actually the space of Jacobi forms of weight=0 and index=1 is known to be 
one-dimensional and thus the above result could have been guessed without explicit computation. We find that $Z_{ell}(K3)(\tau;z=0)=24$ and 
$Z_{ell}(K3)(\tau;z=1/2)=16+{\cal O}(q)$  and thus (\ref{K3-elliptic-genus}) reproduces the Euler number and signature of K3.

In Ref.~\cite{EOTY} and more recently in \cite{EH1} the expansion of the K3 elliptic genus in terms of irreducible representations of ${\cal N}=4$ superconformal algebra has been discussed in detail. We first provide the data of representation theory \cite{ET1,ET2}. 
For a rigorous mathematical exposition, see e.g.~\cite{Kac,KacWakimoto}.

Let us introduce the character formula of 
the BPS (short) representation of spin $\ell=0$ in Ramond sector with $(-1)^F$ insertion
\begin{eqnarray}
&&\ch^{{\hat{R}}}_{h={1\over 4},\ell=0}(\tau;z)={\theta_1(\tau;z)^2\over \eta(\tau)^3}\mu(\tau;z),\\
&&\mu(\tau;z)={-ie^{\pi iz}\over \theta_1(\tau;z)}\sum_{n=-\infty}^{\infty}(-1)^n{q^{{1\over 2}n(n+1)}e^{2\pi inz}\over 1-q^ne^{2\pi iz}}.\label{mu}
\end{eqnarray}
The BPS representation has a non-vanishing index
\begin{equation}
\ch^{{\hat{R}}}_{h={1\over 4},\ell=0}(\tau;z=0)=1.
\end{equation}
We also introduce the character of a non-BPS (long) representation with conformal dimension $h$\begin{equation}
q^{h-{3\over 8}}\,{\theta_1(\tau;z)^2\over \eta(\tau)^3}.
\end{equation}

 Then the elliptic genus is expanded as  
   \begin{eqnarray}
&&Z_{ell}(K3)(\tau;z)=24\ch^{{\hat{R}}}_{h={1\over 4},\ell=0}(\tau;z)
+\Sigma(\tau){\theta_1(\tau;z)^2\over \eta(\tau)^3}\label{decomp1}
\end{eqnarray}
where the expansion function $\Sigma(\tau)$ is given by
\begin{eqnarray}
\Sigma(\tau)&=&-8\left[\mu(\tau;z={1\over 2})+ \mu(\tau;z={1+\tau\over 2})+ 
\mu(\tau;z={\tau\over 2})\right] \label{Sigma}   \\
&=&-2\,q^{-1/8} \,(1-\sum_{n=1}^\infty A_n q^n)
\end{eqnarray}
If one uses the relation that  the non-BPS representation splits into a sum of BPS representations at the unitarity bound  $h=1/4$,
 \begin{eqnarray}
 q^{-{1\over 8}}{\theta_1(\tau;z)^2\over \eta(\tau)^3}=
 2\ch^{{\hat{R}}}_{h={1\over 4},\ell=0}(\tau;z)+\ch^{{\hat{R}}}_{h={1\over 4},\ell={1\over 2}}(\tau;z) ,
 \end{eqnarray} 
 the polar term in $\Sigma$ may be eliminated and 
the above decomposition (\ref{decomp1}) can also be written as 
\begin{eqnarray}
&&Z_{ell}(K3)(\tau;z)=20\ch^{{\hat{R}}}_{h={1\over 4},\ell=0}(\tau;z)
-2\ch^{{\hat{R}}}_{h={1\over 4},\ell={1\over 2}}(\tau;z)\nonumber \\
&&\hskip3cm +2\sum_{n=1}^{\infty}A_n q^{n-{1\over 8}}{\theta_1(\tau;z)^2\over \eta(\tau)^3}.
\end{eqnarray}
The coefficients $A_n$  have been computed explicitly for 
lower orders by expanding the series (\ref{Sigma}),
\begin{equation}
\begin{array}{l|rrrrrrrrrrrrrrrrr}
n& 1&2&3&4&5&6&7&8&9& \cdots \\
\hline
A_n & 45 & 231 & 770 & 2277 & 5796 & 13915 & 30843 & 65550 &132825 & \cdots
\label{coefftable}
\end{array}
\end{equation}
and it was conjectured that they are all positive integers  \cite{Ooguri}. 

On the other hand the asymptotic behavior of $A_n$ at large $n$ has recently been derived using the analogue of the Rademacher expansion of modular forms 
\cite{EH1}
\begin{equation}
A_n\approx {2\over \sqrt{8n-1}}e^{2\pi \sqrt{{1\over 2}(n-{1\over 8})}}.\label{asymptoic}
\end{equation}
It turns out that the above formula (\ref{asymptoic}) gives a good estimate 
of $A_n$ even at smaller values of $n$ and this confirms the positivity of the coefficients $A_n$.
Note that the series $\mu(\tau;z)$ (\ref{mu}) has the form of a 
Lerch sum (or mock theta function) and thus has a complex modular transformation law which involves  Mordell's integral. In such a situation we can use the method recently developed by mathematicians \cite{Zwegers, Bringmann-Ono, Zagier} and construct the Poincar\'e-Maas series to derive the above asymptotic formula. 

The above table contains a surprise: the first 5 coefficients, $A_1,..., A_5$, 
are equal to dimensions of {\it irreducible} 
representations of $M_{24}$, the largest Mathieu group, 
see Appendix~\ref{AppA}. The coefficients $A_6$ and $A_7$ can also be 
nicely decomposed as \begin{align}
A_6  &= 3520 + 10395, \\
A_7 &= 10395 +  5796 + 5544 + 5313 + 2024 + 1771
\end{align}  into the sum of dimensions\footnote{The tentative decomposition of $A_7$ shown in the previous of this paper was not correct in view of the later study of twisted elliptic genus in \cite{Cheng:2010pq,Gaberdiel:2010ch}. Here it is corrected according to their papers.}.
For $n\geq 8$, it is still possible to decompose $A_n$ into a sum of 
dimensions of irreducible representations  of $M_{24}$, but decompositions
are not as unique.\footnote{It may also be interesting to point out that
$2, 3, 5, 7, 11,$ and $23$ appear in prime factorization of $A_n$
more frequently than any other prime numbers and 
with certain periodicities in $n$. These are also prime factors of 
the order of $M_{24}$.}

This observation can be compared to the famous observation of McKay and Thompson\cite{Thompson}: there, the first few terms of the expansion coefficients of $J(q)$, \begin{equation}
J(q)=\frac1q+196884+21493760q^2+\ldots
\end{equation} could be naturally decomposed into the sum of the dimension of the irreducible representation of the monster simple group.
Conway and Norton \cite{ConwayNorton} formulated it  in terms of an infinite dimensional graded representation of the monster group $\bigoplus_i V_i$ such that $\dim V_i$ is the coefficient of $q^i$ of $J(q)$, and called this observation the monstrous moonshine. 
Frenkel, Leopwski and Meurman then found \cite{FLM} that this representation is naturally assocaited to the two-dimensional string propagating on $\mathbb{R}^{26}/\Lambda/\mathbb{Z}_2$ where $\Lambda$ is the Leech lattice. See e.g.~\cite{Gannon} for a recent review.

In our case the existence of a natural vector space whose graded dimension gives $\Sigma(q)$ is guaranteed by construction: 
it is the Hilbert space of the
two-dimensional supersymmetric conformal field theory
whose target space is $K3$.
The problem is to identify the action of $M_{24}$ on it.\footnote{
Dong and Mason pursued the analogue of the monstrous moonshine in the case of $M_{24}$, see e.g.~\cite{DongMason} and references therein. So far there is no direct connection of their work and the geometry of K3.
}

The non-Abelian symplectic symmetry of K3 was studied mathematically by \cite{Mukai,Kondo}. Mukai enumerated eleven K3 surfaces which possess finite non-Abelian automorphism groups. It turns out that all these groups are various subgroups of the Mathieu group $M_{24}$, see Appendix~\ref{AppB} for more details. Is it possible that these automorphism groups at isolated points in the moduli space of K3 surface are enhanced to $M_{24}$ over the whole of moduli space when we consider the elliptic genus? This question is currently under study using Gepner models and matrix factorization.

As discussed in \cite{EH2}, expansion coefficients of 
elliptic genera of hyperK\"ahler manifolds in general have 
an exponential growth and are 
closely related to the black hole entropy. In particular in the case of k-th symmetric product of K3 surfaces we obtain the leading behavior 
\begin{equation}
A_n\approx e^{2\pi \sqrt{{k^2\over k+1}n-\left({k\over 2(k+1)}\right)^2}}
\end{equation} 
which gives the entropy of the standard D1-D5 black hole $S\approx 2\pi \sqrt{kn}$ at large $k$ ($k=Q_1Q_5$ where $Q_1$ and $Q_5$ are the numbers of 
D1 and
D5 branes). Thus the elliptic genus of K3 surface may be considered as describing the multiplicity of microstates of a small black hole with $Q_1=Q_5=1$.

Here the situation is somewhat similar to a model of black hole described by Witten in \cite{Witten}, 
where microstates of a small black hole span the representation space of the monster group. Although the partition function of the theory is discussed, the relevant CFT is modular invariant separately in left and right sectors and the 
discussion is effectively the same as considering the elliptic genus. 

It will be extremely interesting to see if the Mathieu group $M_{24}$ in fact acts on the elliptic genus of K3.

\section*{Acknowledgment}
The authors thank the hospitality of the Aspen Center of 
Physics during the workshop ``Unity of String Theory,'' 
when the observation reported in this paper was made.
We would like to thank S. Mukai for discussions. 

 T. E. is supported in part by Grant in Aid from the Japan Ministry of
Education, Culture, Sports, Science and Technology (MEXT). 
H. O. is supported in part by U.S. Department of Energy 
grant DE-FG03-92-ER40701, 
the World Premier International Research Center Initiative 
and a Grant-in-Aid for Scientific Research (C) 20540256
of MEXT, and the Humboldt Research Award.
Y. T. is supported in part by the NSF grant PHY-0503584, 
and by the Marvin L. Goldberger membership at the Institute for Advanced Study. 
\appendix

\section{Data of $M_{24}$}\label{AppA}
The largest of the Mathieu group, $M_{24}$, has \begin{equation}
2^{10}\cdot 3^3 \cdot 5 \cdot 7 \cdot 11 \cdot 23 = 244823040
\end{equation}
elements.  
There are 26 conjugacy classes and 26 irreducible representations.
The character table is given in Table~\ref{character-table},
whose data is taken from \cite{Iwanami,Atlas}. The conjugacy class is labeled according to the convention of \cite{Atlas}.
In the character table, $e^\pm_p$ stands for \begin{equation}
e^\pm_p=(\pm\sqrt{-p}-1)/2.
\end{equation}

The dimensions of the irreducible representations are, in the increasing order, \begin{align}
&1, 23, 45, 45, 231, 231, 252, 253, 483, 770, 770,\nonumber \\
& 990, 990, 1035, 1035, 1035, 1265, 1771, 2024,\nonumber \\
&  2277, 3312, 3520, 5313, 5796, 5544, 10395.
\end{align} Here the irreducible representations of dimensions \begin{equation}
45, 231, 770, 990, 1035
\end{equation} come in complex conjugate pairs. There is in addition an extra real $1035$-dimensional irreducible representation.

\begin{table}
\vspace*{-4ex}
\centerline{\scalebox{.73}{\rotatebox{270}{$\begin{array}{rrrrrrrrrrrrrrrrrrrrrrrrrrrrrrrrrrrrrr}
\text{1A} & \text{2A} & \text{3A} &\text{5A} & \text{4B} &\text{7A}&\text{7B}& \text{8A} & \text{6A} & \text{11A}&\text{15A} & \text{15B} & \text{14A} & \text{14B} & \text{23A} &\text{23B} & \text{12B}& \text{6B} & \text{4C} & \text{3B} & \text{2B} &\text{10A} & \text{21A} & \text{21B} & \text{4A} &\text{12A} \\
\hline
 1 & 1 & 1 & 1 & 1 & 1 & 1 & 1 & 1 & 1 & 1 & 1 & 1 & 1 & 1 & 1 & 1 & 1 & 1 & 1 & 1 & 1 & 1 & 1 & 1 & 1 \\
 23 & 7 & 5 & 3 & 3 & 2 & 2 & 1 & 1 & 1 & 0 & 0 & 0 & 0 & 0 & 0 & -1 & -1 & -1 & -1 & -1 & -1 & -1 & -1 & -1 &
   -1 \\
 252 & 28 & 9 & 2 & 4 & 0 & 0 & 0 & 1 & -1 & -1 & -1 & 0 & 0 & -1 & -1 & 0 & 0 & 0 & 0 & 12 & 2 & 0 & 0 & 4 & 1
   \\
 253 & 13 & 10 & 3 & 1 & 1 & 1 & -1 & -2 & 0 & 0 & 0 & -1 & -1 & 0 & 0 & 1 & 1 & 1 & 1 & -11 & -1 & 1 & 1 & -3 &
   0 \\
 1771 & -21 & 16 & 1 & -5 & 0 & 0 & -1 & 0 & 0 & 1 & 1 & 0 & 0 & 0 & 0 & -1 & -1 & -1 & 7 & 11 & 1 & 0 & 0 & 3 &
   0 \\
 3520 & 64 & 10 & 0 & 0 & -1 & -1 & 0 & -2 & 0 & 0 & 0 & 1 & 1 & 1 & 1 & 0 & 0 & 0 & -8 & 0 & 0 & -1 & -1 & 0 &
   0 \\
 45 & -3 & 0 & 0 & 1 & e_7^+ & e_7^- & -1 & 0 & 1 & 0 & 0 & -e_7^+ & -e_7^- & -1 & -1 & 1 &
   -1 & 1 & 3 & 5 & 0 & e_7^- & e_7^+ & -3 & 0 \\
 45 & -3 & 0 & 0 & 1 & e_7^- & e_7^+ & -1 & 0 & 1 & 0 & 0 & -e_7^- & -e_7^+ & -1 & -1 & 1 &
   -1 & 1 & 3 & 5 & 0 & e_7^+ & e_7^- & -3 & 0 \\
 990 & -18 & 0 & 0 & 2 & e_7^+ & e_7^- & 0 & 0 & 0 & 0 & 0 & e_7^+ & e_7^- & 1 & 1 & 1 & -1
   & -2 & 3 & -10 & 0 & e_7^- & e_7^+ & 6 & 0 \\
 990 & -18 & 0 & 0 & 2 & e_7^- & e_7^+ & 0 & 0 & 0 & 0 & 0 & e_7^- & e_7^+ & 1 & 1 & 1 & -1
   & -2 & 3 & -10 & 0 & e_7^+ & e_7^- & 6 & 0 \\
 1035 & -21 & 0 & 0 & 3 & 2 e_7^+ & 2 e_7^- & -1 & 0 & 1 & 0 & 0 & 0 & 0 & 0 & 0 & -1 & 1 & -1 & -3 &
   -5 & 0 & -e_7^- & -e_7^+ & 3 & 0 \\
 1035 & -21 & 0 & 0 & 3 & 2 e_7^- & 2 e_7^+ & -1 & 0 & 1 & 0 & 0 & 0 & 0 & 0 & 0 & -1 & 1 & -1 & -3 &
   -5 & 0 & -e_7^+ & -e_7^- & 3 & 0 \\
 1035 & 27 & 0 & 0 & -1 & -1 & -1 & 1 & 0 & 1 & 0 & 0 & -1 & -1 & 0 & 0 & 0 & 2 & 3 & 6 & 35 & 0 & -1 & -1 & 3 &
   0 \\
 231 & 7 & -3 & 1 & -1 & 0 & 0 & -1 & 1 & 0 & e_{15}^+ & e_{15}^- & 0 & 0 & 1 & 1 & 0 & 0 & 3 & 0 & -9 & 1 &
   0 & 0 & -1 & -1 \\
 231 & 7 & -3 & 1 & -1 & 0 & 0 & -1 & 1 & 0 & e_{15}^- & e_{15}^+ & 0 & 0 & 1 & 1 & 0 & 0 & 3 & 0 & -9 & 1 &
   0 & 0 & -1 & -1 \\
 770 & -14 & 5 & 0 & -2 & 0 & 0 & 0 & 1 & 0 & 0 & 0 & 0 & 0 & e_{23}^+ & e_{23}^- & 1 & 1 & -2 & -7 & 10 & 0
   & 0 & 0 & 2 & -1 \\
 770 & -14 & 5 & 0 & -2 & 0 & 0 & 0 & 1 & 0 & 0 & 0 & 0 & 0 & e_{23}^- & e_{23}^+ & 1 & 1 & -2 & -7 & 10 & 0
   & 0 & 0 & 2 & -1 \\
 483 & 35 & 6 & -2 & 3 & 0 & 0 & -1 & 2 & -1 & 1 & 1 & 0 & 0 & 0 & 0 & 0 & 0 & 3 & 0 & 3 & -2 & 0 & 0 & 3 & 0 \\
 1265 & 49 & 5 & 0 & 1 & -2 & -2 & 1 & 1 & 0 & 0 & 0 & 0 & 0 & 0 & 0 & 0 & 0 & -3 & 8 & -15 & 0 & 1 & 1 & -7 &
   -1 \\
 2024 & 8 & -1 & -1 & 0 & 1 & 1 & 0 & -1 & 0 & -1 & -1 & 1 & 1 & 0 & 0 & 0 & 0 & 0 & 8 & 24 & -1 & 1 & 1 & 8 &
   -1 \\
 2277 & 21 & 0 & -3 & 1 & 2 & 2 & -1 & 0 & 0 & 0 & 0 & 0 & 0 & 0 & 0 & 0 & 2 & -3 & 6 & -19 & 1 & -1 & -1 & -3 &
   0 \\
 3312 & 48 & 0 & -3 & 0 & 1 & 1 & 0 & 0 & 1 & 0 & 0 & -1 & -1 & 0 & 0 & 0 & -2 & 0 & -6 & 16 & 1 & 1 & 1 & 0 & 0
   \\
 5313 & 49 & -15 & 3 & -3 & 0 & 0 & -1 & 1 & 0 & 0 & 0 & 0 & 0 & 0 & 0 & 0 & 0 & -3 & 0 & 9 & -1 & 0 & 0 & 1 & 1
   \\
 5796 & -28 & -9 & 1 & 4 & 0 & 0 & 0 & -1 & -1 & 1 & 1 & 0 & 0 & 0 & 0 & 0 & 0 & 0 & 0 & 36 & 1 & 0 & 0 & -4 &
   -1 \\
 5544 & -56 & 9 & -1 & 0 & 0 & 0 & 0 & 1 & 0 & -1 & -1 & 0 & 0 & 1 & 1 & 0 & 0 & 0 & 0 & 24 & -1 & 0 & 0 & -8 &
   1 \\
 10395 & -21 & 0 & 0 & -1 & 0 & 0 & 1 & 0 & 0 & 0 & 0 & 0 & 0 & -1 & -1 & 0 & 0 & 3 & 0 & -45 & 0 & 0 & 0 & 3 &
   0
\end{array}$}}}
\caption{Character table of $M_{24}$. \label{character-table}}
\end{table}

\newpage

\section{$M_{24}$ and the classical geometry of K3}\label{AppB}
Here we briefly summarize the relation between the classical geometry of the K3 surface and $M_{24}$, first found in \cite{Mukai} and elaborated in \cite{Kondo}.

Before proceeding, we need to recall the definition of $M_{24}$.
Of many equivalent ways to define it, one that is most understandable to string theorists is to use an even self-dual lattice of dimension 24.
Consider the root lattice of $A_1$ whose generator has squared length $2$.  
Let us denote its weight lattice by $A_1^*$ whose generator has squared length $1/2$.
Take the 24-dimensional lattice $A_1{}^{24}$. This is even but not self-dual, because the dual lattice is $A_1^*{}^{24}$. An even self-dual lattice $N$ containing $A_1{}^{24}$ will have the structure \begin{equation}
A_1{}^{24} \subset N \subset A_1^*{}^{24}.
\end{equation} Let $\GG=N/A_1{}^{24}$, which is a vector subspace
of $A_1^*{}^{24}/A_1{}^{24}\simeq \bZ_2{}^{24}$.
Let us represent an element of $\GG$ by a sequence of twenty-four 0 and 1, and define the weight of a vector to be the number of 1's in it.

The self-duality of $N$ translates to the fact $\GG$ is 12-dimensional.
The evenness translates to the fact that the weight of every element of $\GG$ is a multiple of 4. Let us further demand the vectors of $N$ whose length squared are two are the roots of $A_1{}^{24}$ and not more. Then $\GG$ does not have an element with weight 4. 

These conditions fix the form of $\GG$ uniquely, and $\GG$ is known as the  extended binary Golay code. $M_{24}$ is defined as the subgroup of the permutation $S_{24}$ of the coordinates of $\bZ_2{}^{24}$ which preserves $\GG$. 

The lattice $N$ thus constructed defines a chiral CFT with $c=24$ whose current algebra is $A_1{}^{24}$. Therefore $M_{24}$ is the discrete symmetry of this chiral CFT.

Now let us recall that the cohomology lattice of K3, \begin{equation}
\Lambda=H^*(K3,\bZ)
\end{equation} is also an even self-dual lattice of dimension 24,
but with signature $(4,20)$. The close connection between $M_{24}$ and the geometry of the K3 surface stems from this fact.

Take a K3 surface $S$, and let $G$ its symmetry preserving the holomorphic 2-form. Let $\Lambda^G$ the part of $\Lambda$ preserved by $G$, and $\Lambda_G$ its orthogonal complement.
$\Lambda_G$ is inside the primitive part of $H^{1,1}$, thus it is negative definite.  Using Nikulin's result, it can be shown that $\Lambda_G$ is a sublattice of $N$. Therefore $G$ is a subgroup of $M_{24}$. 

$G$ cannot be $M_{24}$ itself, however. The action of $G$ on $S$ preserves at least $H^0$, $H^4$, $H^{2,0}$, $H^{0,2}$ and the K\"ahler form. 
Hence $\Lambda^G$ is at least five-dimensional, and $\Lambda_G$ is at most 19-dimensional. 
This implies that the action of $G$ on $N$ as real linear maps should at least have five-dimensional fixed subspace. 
This translates to the fact that the action of $G$ on 24 points as a subgroup of $M_{24}$ splits them into at least five orbits. 

Similarly, starting from a subgroup $G$ of $M_{24}$ which acts on 24 points with at least five orbits, one can construct the action of $G$ on $H^{1,1}$. Using the global Torelli theorem, this translates to the existence of a K3 surface $S$ whose symmetry is $G$.

One example is the Fermat quartic, $X^4+Y^4+Z^4+W^4=0$ in $\mathbb{CP}^3$. The symmetry is $(\bZ_4)^2 \rtimes S_4$, with 384 elements.  This is indeed a subgroup of $M_{24}$ which decomposes 24 points into five orbits, of length 1, 1, 2, 4 and 16. 

More examples and details of the analysis can be found in \cite{Mukai} and in \cite{Kondo}.

\end{document}